\documentclass[12pt]{article}
\usepackage{epsfig}
\begin{document}

\begin {center}
{\bf An alternative fit to Belle mass spectra for  $D\bar D$, $D^*\bar
D^*$ and $\Lambda _C \bar \Lambda _C$}

\vskip 4mm {D.\ V.\ Bugg\footnote{email:
david.bugg@stfc.ac.uk} \\[2mm]
{\normalsize\it Queen Mary, University of
London, London E1\,4NS, UK} \\[3mm]}
\end {center}
\date{\today}

\begin{abstract}
\noindent
Peaks observed by Belle in $D\bar D$ at 3.878 GeV and in $D^*\bar D^*$
at 4.156 GeV may be fitted by phase space multiplied by a form
factor with an RMS radius of interaction 0.63 fm.
The peak observed in  $\Lambda _C\bar \Lambda _C$ at 4.63 GeV
may be explained by $Y(4660)$, multiplied by a corresponding form factor
with RMS radius $\sim 0.94$ fm.

\vskip 2mm

\end{abstract}
\vskip 4mm

Belle find a broad bump in $D\bar D$ with mass $M=3878 \pm 48$
MeV, $\Gamma = 347 ^{+316}_{-143}$ MeV \cite {BelleDD}.
The data are reproduced in Fig. 1(a) after a subtraction of
experimental background.
Belle tentatively interpret this as a broad resonance denoted
$X(3880)$ with $3.8\sigma$ significance. They conclude that 
'the observed threshold enhancement is not consistent with 
non-resonant $e^+e^- \to J/\Psi D\bar D$.'

\begin{figure}[htb]
\begin{center}
\vskip -12mm
\epsfig{file=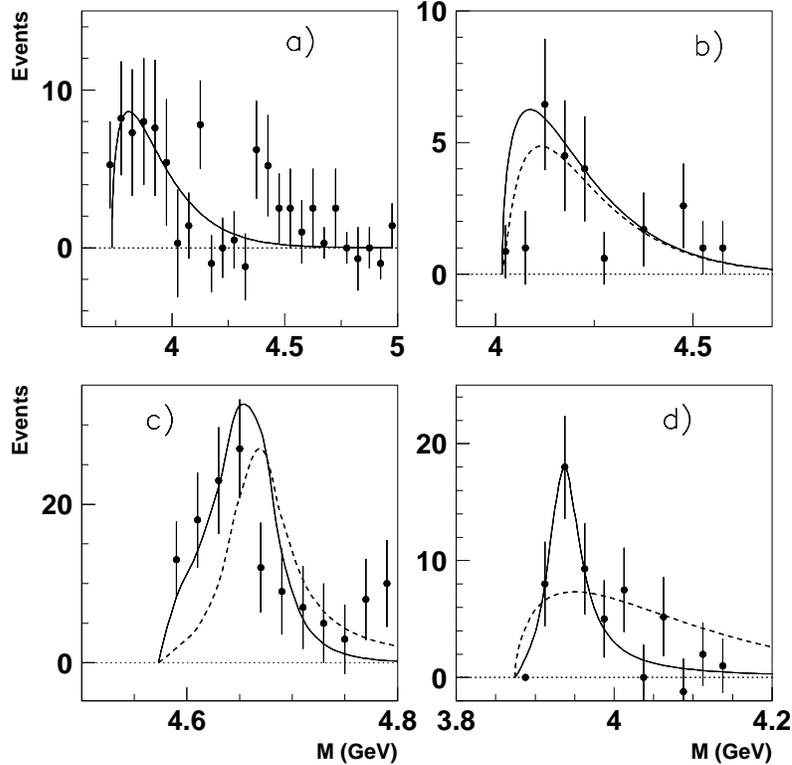,width=12cm}
\vskip -6mm
\caption
{Fits to Belle data on (a) $D\bar D$, (b) $D^*\bar D^*$, (c) $\Lambda
_C \bar \Lambda _C$ and (d) $D\bar D^*$.
In (b), the full curve is for S-wave $D^*\bar D^*$ and the dashed
curve shows the perturbation due to a P-state centrifugal barrier.
In (c), the full curve shows the line-shape of $Y(4663)$ after
modulation by a form factor; the dashed curve is the result without the
form factor.
In (d), the full curve shows the Belle fit with $Y(3942)$
and the dashed curve the fit with phase space and a form factor.}
\end{center}
\end{figure}

The proposal made here is that this spectrum has an intensity
proportional to $D\bar D$ phase space $\rho (s)$ multiplied by the
square of a form factor $\exp (-q^2 R^2/6)$ for a Gaussian source;
$q$ is the momentum of $D$ and $\bar D$ in their centre of mass
and $R$ is the radius of interaction of the $D\bar D $ pair:
\begin {eqnarray}
I(s) &=& \rho (s) e^{-2Aq^2} , \\
A    &=& \frac {1}{6}\left( \frac {R(fm)}{\hbar c} \right)^2,
\end {eqnarray}
with $\hbar c = 0.19732$ GeV/c.
The assumption being made here is that the final-state
interaction is fairly weak and that the amplitude may be
parametrised by a scattering length, with the exponential
providing an effective range.
No resonance is involved and the data are not particularly sensitive to small
phase shifts.

The full curve in Fig. 1(a) uses $A = 1.7$ (GeV/c)$ ^{-2}$,
corresponding to a reasonable RMS radius $R = 0.63$ fm for the combined
$D\bar D$ pair or $R' = R/\sqrt {2} = 0.45$ fm for each $D$.
Although the fit looks slightly ragged, it is in fact very close
to that made by Belle with a broad resonance.
It may appear surprising that the form factor has a strong effect;
the reason is that momenta increase rapidly from threshold because
of the high masses of the two $D$.

Secondly, Belle also observe a peak in $D^*\bar D^*$ with
$M = 4156 ^{+25}_{-20}(stat) \pm 15(syst)$ MeV, $\Gamma = 139
^{+111}_{-61} \pm 21$ MeV, reproduced in Fig. 1(b) after
subtracting a very small experimental background. The full curve
shows a fit using Eqs. (1) and (2) with exactly the same radius parameter 
as for Fig. 1(a).
There is some scatter in experimental points, but the fit it reasonable
in view of present statistics.
Belle point out that the peak of Fig. 1(b) is too strong to be explained
by $\Psi (4160)$, for which $<1$ event is to be expected.
One should also note that $\Psi (4160)$ is observed in five other sets of
data \cite {BES2}, \cite {Pakh2}, \cite {Cleo}, in all of which there are 
strong interferences with $\Psi (4040)$ and $X(4260)$, which has $J^{PC} 1^{--}$.
Belle say: 'We interpret the observed enhancement, which has a statistical
significance of $5.5\sigma$, as a new resonance and denote it as $X(4160)$',
i.e. distinct from $\Psi (4160)$.

There are two further sets of data where a relation between them can be
explained by a form factor related to that given above.
Firstly Belle report a sharp peak in
$e^+e^- \to \gamma _{ISR}\Psi '(3686)\pi \pi$
with $M = 4664 \pm 11 \pm 5$ MeV, $\Gamma = 48 \pm 15 \pm 3$ MeV
\cite {Y4664}.
As Guo, Hanhart and Meissner point out \cite {Hanhart},
this coincides with the sharp $\Psi '(3686)f_0(980)$ threshold.
Guo et al. favour interpretation as a dynamically generated molecular
state.
There is a well established mechanism by which a shape threshold can
generate a resonance or attract a pre-existing state \cite {Sync}.
An alternative explanation of $Y(4664)$ is the $\Psi (5S)$ state
or $\Psi (3D)$ \cite {Ding}, though it is then rather narrow.

Secondly, Belle data on $e^+e^- \to \gamma _{ISR}\Lambda _C \bar \Lambda _C$
reveal a narrow peak with
$M=4634 {^{+8}_{-7}} {^{+5}_{-8}}$ MeV,
$\Gamma = 92 {^{+40}_{-24}} {^{+10}_{-21}}$ MeV \cite {LLC}.
These data are reproduced in Fig. 1(c). 
The dashed curve shows a fit using parameters of $Y(4664)$, except
that the width is increased by one standard deviation.
Clearly the dashed curve disagrees with the data.

It seems likely that $Y(4664)$ and $Y(4634)$ are related.
What happens if a form factor is introduced?
For $Y(4664)$, there is rather little effect, since the signal
is centred at the threshold for $\Psi ' f_0(980)$.
However, for $Y(4634)$ some effect is to be expected.
Because the $\Lambda _C$ contains three quarks, one expects the radius
of interaction in this case to be larger than for
$D\bar D$ and $D^*\bar D^*$.
It is well known that the total cross section for $NN$ is
asymptotically larger than for $\pi N$ by a factor $\sim 1.5$.
This increase arises from changing 2 quarks to three in one
particle.
For $\Lambda _C \bar \Lambda _C$ both particles contain 3 quarks.
The full curve of Fig. 1(c) shows a fit assuming $R^2$ increases
between $D\bar D$ and $\Lambda _C\bar \Lambda _C$ by
a factor $(1.5)^2$.
This curve approximately reproduces the peak mass in $\Lambda _C
\bar \Lambda _C$ and also the increase in width.
To achieve this result, it is necessary to increase the width of
$Y(4664)$ by one standard deviation.
A further possible source of a large radius of interaction is that
it is well known that $p\bar p$ and $K^-p$ total cross sections
increase rapidly near threshold. 
A similar effect for $\Lambda _C\bar \Lambda _C$ would account
for the threshold peak in that channel.

A final point concerns Belle data for $D\bar D^*$ \cite {X3943}.
The full curve of Fig. 1(d) shows their fit.
The data require $M = 3942 ^{+7}_{-6} \pm 6$ MeV,
$\Gamma = 37 ^{+26} _{-18} \pm 8$ MeV.
In this case, the fit with phase space and a simple form factor
(shown by the dashed curve) does {\it not} reproduce the data
accurately.
So this does look like a resonance.
Confirmation of this peak in $D\bar D^*$ and its spin-parity is important.

In conclusion, experimentalists and phenomenologists should keep a
watchful eye open for simple non-resonant explanations of bumps
in data.
A form factor of reasonable radius of interaction can produce
two of the peaks reported by Belle and provide an explanation of
the shift of mass between peaks they observe in $\Psi 'f0(980)$ and
$\Lambda _C \bar \Lambda _C$.

\end{document}